\documentstyle[12pt,epsf,cite]{article}

\def\reff#1{(\ref{#1})}

\newcommand{\C}{{\bf C}}
\newcommand{\R}{{\bf R}}
\newcommand{\Z}{{\bf Z}}
\newcommand{\D}[1]{D^{\omega}(#1)}
\renewcommand{\c}[3]{C_{#1#2}{}^{#3}}
\renewcommand{\d}[3]{\Delta_{#1}{}^{#2#3}}
\newcommand{\s}[2]{S^{#1}{}_{#2}}
\newcommand{\zcfs}{Z^{\mbox{\scriptsize CFS}}}

\newcommand{\zdw}{Z^{\mbox{\scriptsize DW}}}
\newcommand{\elle}[2]{{}^#1 \rule{0.3pt}{2.5ex}\rule{1em}{0.3pt}_{\kern
  -10pt  \raisebox{-8pt}{${}^#2$}}}

\def\abstract#1{\begin{center}{\large ABSTRACT}\end{center} \par #1}
\def\title#1{\begin{center}{\Large\bf #1}\end{center}}
\def\author#1{\begin{center}{\sc #1}\end{center}}
\def\address#1{\begin{center}{\it #1}\end{center}}

\begin{document}
\begin{titlepage}
\hspace*{\fill}
\vbox{
  \hbox{TIT/HEP-255}
  \hbox{hep-th/9405099}
  \hbox{April 1994}}
\vspace{7mm}
\title{
  On  three-dimensional topological field theories \\
  constructed from $D^\omega(G)$ for finite groups}
\vskip 1cm
\author{
  Masako ASANO \footnote{E-mail address: \tt maa@phys.titech.ac.jp} and
  Saburo HIGUCHI%
  \footnote{\parbox[t]{15cm}{%
      Fellow of the Japan Society for the Promotion of Science\\
      E-mail address: {\tt hig@phys.titech.ac.jp}}}}
\address{
  Department of Physics, \\
  Tokyo Institute of Technology, \\
  Oh-Okayama, Meguro,\\
  Tokyo 152, Japan}
\vspace{\fill}
\abstract{
  We investigate the 3d lattice topological field theories
  defined by Chung, Fukuma and  Shapere. 
  We concentrate on  the model defined 
  by taking a  deformation $\D{G}$ of the quantum double of a 
  finite commutative group $G$ as the underlying Hopf algebra.
  It is suggested that 
  Chung-Fukuma-Shapere partition function is related to that of
  Dijkgraaf-Witten  by 
  $\zcfs = |\zdw|^2$ when $G=\Z_{2N+1}$.
  For $G=\Z_{2N}$, such a relation does not hold.
}
\vspace{\fill}
\end{titlepage}

\section{Introduction}
Three-dimensional topological field theories have been
attracting interests of mathematicians and physicists. 
There are several ways of constructing 3d topological field theories
\cite{wi,dw,tv,bou2,ac}. 
It is important to study a relation among them
in view of the ultimate goal of classifying all topological
field theories. 
Recently Chung, Fukuma and Shapere 
defined a class of 3d lattice  topological field theories \cite{cfs}.
They are in  one-to-one correspondence with
`nice' Hopf algebras satisfying a few conditions.
It is interesting that the class of Hopf algebras is not
identical with the class of ribbon Hopf algebras, on which 
Chern-Simons type theories are based \cite{rt}.
So we are interested in establishing a relation between
Chern-Simons theory and Chung-Fukuma-Shapere (CFS) theory. 

In this note, We would like to investigate  relations between 
CFS theory for finite groups and
Chern-Simons theory for finite gauge groups (Dijkgraaf-Witten
theory \cite{dw}).
We define CFS theory for finite groups 
by taking $\D{G}$, a deformation of the quantum double of a finite
commutative group $G$, as the underlying Hopf algebra.
It is worth noting that, in every known construction
of 3d topological filed theories, models based on finite groups
have rich structure  and are suitable for explicit calculations
\cite{dw,ac,bou}.   
In the present case, the theory becomes completely rigorous and
is free from divergences because the Hopf algebra is finite-dimensional.

We find that CFS partition function is related to
Dijkgraaf-Witten (DW) partition function by
\begin{equation}
\zcfs_{\D{G}}(M) = |\zdw_{(G;\omega)}(M)|^2
\end{equation}
if one of the following conditions is met: 
(i) $\omega\equiv 1$. (ii) $G=\Z_N$ with odd $N$ and the 3-manifold
$M$ can be constructed from the 3-sphere , lens spaces, or 
  $2\mbox{-manifold} \times S^1$ via the connected sum.
We also find that some CFS theories constructed from $D^\omega(\Z_{2N})$
do not fall into the category of DW theories for finite cyclic groups.

\section{Chung-Fukuma-Shapere theory}
Let us recall the definition of  CFS theory.
We  pick a semi-simple Hopf algebra
$(A;m,u,\Delta,\epsilon,S)$ over $\C$. 
Symbols $m$, $u$, $\Delta$, $\epsilon$ and $S$ denote
multiplication, unit, comultiplication, counit and antipode, respectively. 
Using a basis $\{\phi_x|x \in X\}$ of $A$, we write these operations as
\begin{eqnarray}
  m(\phi_x\otimes\phi_y) & = & \c{x}{y}{z}\phi_z ,\\
  u(1) & = & u^x\phi_x,\\
  \Delta(\phi_x) & = & \d{x}{y}{z}\phi_y\otimes\phi_z,\\
  \epsilon(\phi_x) & = & \epsilon_x ,\\
  S(\phi_x) & = & \s yx \phi_y,
\end{eqnarray}
where $\c{x}{y}{z}, u^x, \epsilon_x, 1 \in \C$.
Summation over the repeated indices is assumed hereafter.
We define the metric $g_{xy}$ and the cometric $h^{xy}$ by
\begin{equation}
  g_{xy}\equiv \c xuv \c yvu ,\qquad h^{xy}\equiv \d uvx \d vuy.
\end{equation}
Since the Hopf algebra $A$ is semi-simple, the existence of the
inverse $g^{xy}$ of $g_{xy}$ is guaranteed.

Furthermore we impose two conditions on the Hopf algebra $A$:
\begin{eqnarray}
  & & h^{xy}  \mbox{\ has the inverse\  }h_{xy},
  \label{invertibleh}\\
  & & h^{xz}g_{zy}=\Lambda \s xy \quad
  (\Lambda \equiv \epsilon^x u_x  = |X|),
  \label{involutives}
\end{eqnarray}
where $|X|$ denotes the order of the set $X$.
In virtue of (\ref{invertibleh}),
we raise and lower the indices $x,y,z,\ldots$ by $g_{xy}$ and $g^{xy}$ 
for $\c xyz$ and $u^x$,
and $h^{xy}$ and $h_{xy}$ for $\d xyz$ and $\epsilon_x$.
Imposing the strong constraint (\ref{involutives}) amounts to requiring that 
applying the direction changing operator (defined below) twice 
is equivalent to the identity operation.

One can calculate the partition function $\zcfs_A(M)$ for a
3-manifold $M$ following the prescription below.
\begin{enumerate}
\item
  Choose an arbitrary lattice $L$ of $M$.
  A lattice $L$ means a composition of polygonal faces which are
  glued together along edges. 
  We assume that every edge in $L$ is a boundary of at least three polygons.
  A simplicial decomposition and its dual are the lattices.
  The number of $i$-cells in $L$ is denoted by $N_i$.
\item
  Decompose $L$ into the set of polygonal faces $F=\{f\}$
  and that of hinges $H=\{h\}$ as depicted in fig.~\ref{fahi}.
\begin{figure}[tbhp]
    \begin{center}
      \leavevmode
       \epsfxsize = 400pt 
       \epsfbox{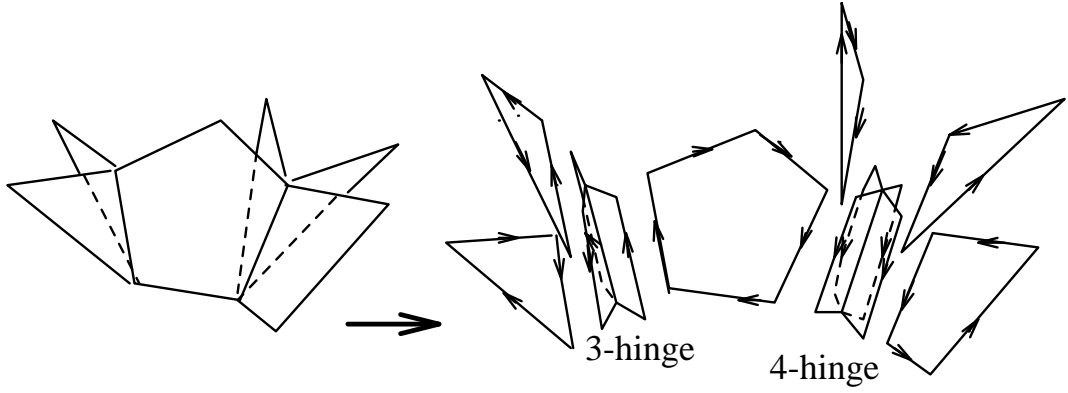}
    \end{center}
    \caption{The decomposition into faces and hinges.
      An $n$-hinge pastes $n$ different faces.}     
\label{fahi}
\end{figure}
  We pick an orientation of each face $f$ and put arrows on an 
  edge according to it.  The edges of an $n$-gon $f$
  is numbered $i=1,\ldots,n (=n_f) $ in this order.
\item
Assign
  \begin{equation}
    C_{x_{f[1]}x_{f[2]}\ldots x_{f[n]}} \equiv
    \c{a_1}{x_{f[1]}}{a_2} \c{a_2}{x_{f[2]}}{a_3} \times \cdots
    \times \c{a_{n-1}}{x_{f[n-1]}}{a_n} \c{a_n}{x_{f[n]}}{a_1} \in \C
  \end{equation}
  to each $n$-gonal face $f$.
  Symbol $f[i]$ stands for one of the integers
  $\{1,\ldots,\sum_{f\in F} n_f \}$
  and $f[i]\ne f'[i']$ if $f\ne f'$ or $i \ne i'$.
  The index $x_i$ runs over $X$.
  One can imagine the variable $x_{f[i]}$ lives on the $i$-th
    edge of $f$.
\item
  The assignment of  an arrow to each edge   of faces induces
  those to $n$-hinges $\{h\}$. 
  The $n$ arrows on the edges of a hinge $h$ are not always in
  the same direction. 

  If all arrows in the $n$-hinge $h$ are in the same direction,
  we number the edges $i=1,\ldots,n (=n_h)$ in the clockwise
  order around the arrows.  
  Let the symbol $h[j]=f[k]$ if and only if the $j$-th edge of $h$ and the
  $k$-th edge of $f$ is glued.
  We associate
  \begin{equation}
    \Delta^{x_{h[1]}x_{h[2]}\cdots x_{h[n]}}
    \equiv  \d {a_1}{x_{h[1]}}{a_2}\d
    {a_2}{x_{h[2]}}{a_3}\times\cdots\times  
    \d {a_{n-1}}{x_{h[n-1]}}{a_{n}}
    \d {a_{n}}{x_{h[n]}}{a_1} \in \C  
    \label{den}
  \end{equation}
  to $h$.  

  If the directions of arrows in $h$ are not the same as the rest,
  we change the direction of the arrows so as to make directions of
  all the arrows match by multiplying an
  additional factor (the direction changing operators) 
  $\s{x}{x'} \in \C $ (See fig.~\ref{dich}) for each hinge. 
\begin{figure}[tbhp]
    \begin{center}
      \leavevmode
      \epsfbox{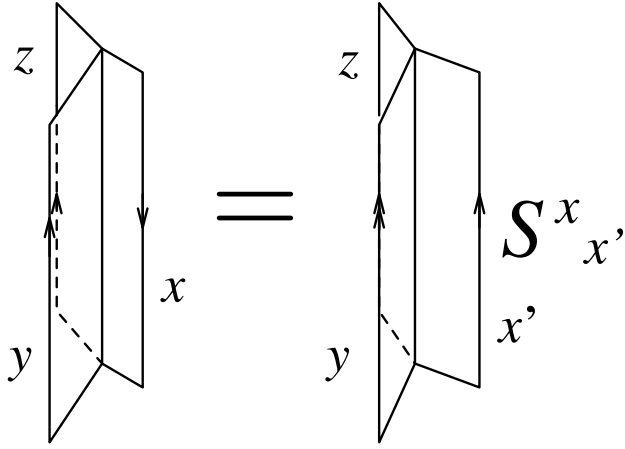}
    \end{center}
    \caption{direction changing operator} 
    \label{dich}
\end{figure}
\item
  We have defined the weight 
  $C_{x_{f[1]}\cdots x_{f[n]}}$ for each face $f$ and 
  $\Delta^{x_{h[1]}\cdots x'_{h[j]} \cdots x_{h[n]}} 
  \prod_{j\in R_h} \s{x_{h[j]}}{x'_{h[j]}}$
  for each hinge $h$. 
  The set $R_h$ corresponds to the set of all direction changing
  edges of a hinge $h$.
  The partition function is defined by contracting indices:
  \begin{equation}
    \zcfs_{A}(M)
    =
    {\cal N} 
    \prod_{f\in F}  C_{x_{f[1]}\cdots x_{f[n_f]}}
    \prod_{h\in H} 
    \left[\Delta^{x_{h[1]}\cdots x'_{h[j]} \cdots x_{h[n_h]}}
    \prod_{j \in R_h} \s{x_{h[j]}}{x'_{h[j]}} \right],
    \label{zfuk}
  \end{equation}
  where $\cal N$ is a normalization factor.

\end{enumerate}
In ref.~\cite{cfs} it is shown that, with an appropriate
normalization, $\zcfs_A(M)$ is a topological invariant. 

\section{Finite-dimensional Hopf Algebras $A$}
As the authors of ref.~\cite{cfs} pointed out,
the partition function (\ref{zfuk}) can suffer from the divergence
in  the normalization factor.
Therefore the definition (\ref{zfuk}) sometimes stays at the formal level.
On the other hand, if we consider a finite-dimensional Hopf algebra $A$,
the definition is completely rigorous.
We will study the case hereafter.

We find that the partition function (\ref{zfuk}) becomes 
topological invariant if we choose the correct normalization 
\begin{equation}
  {\cal N}=\Lambda^{-N_3- N_1} \label{normalization}
\end{equation}
for any finite-dimensional Hopf algebra $A$ %
\footnote{
  A normalization ${\cal N}=\Lambda^{-N_3}$ is employed  for
  $\C[G]$ in ref.~\cite{cfs}. 
  There is, however, no contradiction. We use $\sum_{x\in X}$ to
  sum over indices instead of the Haar measure $\int_G dg$ in ref.\cite{cfs}.
  So an additional factor $\Lambda^{-N_1}$ appears here.
}.

We claim that the partition function for the sphere depends
only on the dimension of the Hopf algebra $A$:
\begin{equation}
  \zcfs_A(S^3)= \Lambda^{-1}=|X|^{-1}. \label{spherepart}
\end{equation}
This is shown as follows. Take, for instance,  a lattice consisting of
three triangular faces and three 3-hinges each of which pastes
the three faces ($N_0=N_1=N_2=N_3=3$).
The partition function is 
$\Lambda^{-6} C_{rst}C_{uvw}C_{xyz} \Delta^{rux} \Delta^{svy} \Delta^{twz}$.
This expression can be reduced to eq.(\ref{spherepart}) as a
consequence of the axioms of Hopf algebras and
eqs.~(\ref{invertibleh}) and (\ref{involutives}). 

\section{$A=\C[G]$ for a finite group $G$}
Let $G$ be a finite group with the unit element $e$.
The group algebra $ \C[G]= \bigoplus_{x\in G}\C \phi_x $,
where $\phi_x$ is a formal basis,
has a natural Hopf algebra structure. 
This finite-dimensional Hopf algebra satisfies the conditions
(\ref{invertibleh}) and (\ref{involutives}) 
and therefore induces a topological field theory \cite{cfs}. 
The partition function becomes
\begin{eqnarray}
  \zcfs_{\C[G]}(M) &=& 
  |G|^{-N_3-N_1 }
  \prod_{f\in F}
  |G|^{N_2}\delta_{x_{f[1]}x_{f[2]}\cdots x_{f[n_f]},e}
  \times \nonumber \\
  & &  \times
  \prod_{h\in H}\left[
  \delta^{x_{h[1]},x_{h[2]}}\delta^{x_{h[1]},x_{h[3]}}\cdots
  \delta^{{x_{h[1]}},x'_{h[j]}}\cdots  \delta^{x_{h[1]},x_{h[k]}} 
  \prod_{j \in R_h} 
  \delta_{x_jx'_j,e}\right]  \nonumber \\
  &=&  |G|^{-N_0}
  \sum_{ \{g([PQ])\}_{[PQ] \in H}}
  \prod_{f\in F}
  \delta_{g(\partial f),e}.
  \label{zlg}
\end{eqnarray}
In the second line,  $g([PQ]) $ running over the group $G$ is the link
variable on the hinge $[PQ]$ between two  vertices $P$ and $Q$.
Note that $g({\partial f})=\prod_{[PQ]\in f} g([PQ])$
and $g([PQ])=g([QP])^{-1}$.
We have used that $\sum_{i=0}^3 (-1)^i N_i = 0$ for closed 3-manifolds.
The summation in eq.(\ref{zlg}) counts the number of flat gauge field
configurations (not up to gauge transformation) 
in the lattice gauge theoretic picture.
The factor $|G|^{-N_0+1}$ cancels the gauge volume of the local
gauge transformations defined on vertices and we have
\begin{equation}
  \zcfs_{\C[G]}(M)= |G|^{-1} |\hom(\pi_1(M), G)|.
  \label{cgpart}
\end{equation}
The partition function \reff{cgpart} is sensitive only of the 
fundamental group of $M$. 
It is not surprising that eq.(\ref{zlg})
catches the information of the fundamental group of $M$. 
CFS theory depends only on the 2-skeleton $L_2$ of the lattice but
$\pi_1(L_2)$ is isomorphic to $\pi_1(M)$.

We note that (\ref{cgpart}) is not proportional to the number
of flat gauge fields up to gauge transformation since we do not
introduce the equivalence relation 
$\exists g \in G$ s.t. $\rho(\cdot) \sim g \rho(\cdot) g^{-1}$ for
$\rho \in \hom(\pi_1(M), G)$.

The partition function (\ref{cgpart}) is exactly the same form
as that of Dijkgraaf-Witten theory 
for the trivial 3-cocycle $\omega \equiv 1$ \cite{dw}.
A 3-cocycle is a map $\omega : G \times G \times G \rightarrow U(1)$
which satisfies the condition
\begin{eqnarray}
& &  \omega(g,x,y)\omega(gx,y,z)^{-1}\omega(g,xy,z)
  \omega(g,x,yz)^{-1}\omega(x,y,z) =1,  \\ 
& & \omega(e,y,z) = \omega(x,e,z) = \omega(x,y,e) =1
\label{eq:cocycle}
\end{eqnarray}
for any $g,x,y,z\in G$. $\omega\equiv 1$ is a 3-cocycle.
In ref.~\cite{dw}, it is argued that the 
Chern-Simons theories with a finite gauge group (DW theories) are labeled by
its 3-cocycles. The partition function of  DW
theory is
\begin{equation}
  \zdw_{(G;\omega)} = \sum_{f\in \hom(\pi_1(M),G)} W(f;\omega),
\end{equation}
where $W(f;\omega)$ is the weight satisfying
$W(f;\omega) \equiv 1$ for $\omega\equiv 1$.

\section{$A=\D{G}$ for a finite commutative group $G$}
In DW theories, the $\omega\not\equiv 1$ theories are more
interesting than the $\omega\equiv 1$  one.
The former theories can distinguish
distinct 3-manifolds  with identical fundamental groups.
Therefore it is natural to ask whether one can obtain DW theory with
non-trivial $\omega$ from CFS theory by  a choice of Hopf algebra $A$.

In the following, we investigate the theory defined by $A=\D{G}$.
$\D{G}$ is a quasi-Hopf algebra \cite{drin} introduced in ref.\cite{rpd}.
This choice seems promising 
since  DW theory from a cocycle $\omega$
 is suggested to be equivalent to the Altschuler-Coste theory
which uses the regular representation of $\D{G}$ \cite{ac}. 

Of course, the quasi-Hopf algebra $\D{G}$ is not always a Hopf algebra.
If $G$ is commutative, however,
it can be verified that $\D G$  becomes a Hopf algebra
and satisfies the conditions (\ref{invertibleh}) and (\ref{involutives}).
{}From now on, we restrict ourselves to the case of commutative
group $G$. 
Let us recall the definition of $\D{G}$ for a commutative finite
group $G$.
$\D{G}$ is spanned by the  formal basis 
$\{\phi_{(g,x)}| g,x \in G \}$
\footnote{The base $\phi_{(g,x)} $is usually written as
  $\elle{g}{x}$\ \ .}
as a $\C$-module.
Hopf algebra structure of $\D{G}$ is 
\begin{eqnarray}
  \c{(g,x)}{(h,y)}{(k,z)} 
  & =  & \delta_{g,h} \delta_{g,k} \delta_{xy,z}\theta_g(x,y), 
  \nonumber \\
  u^{(g,x)} & = & \delta_{x,e},\nonumber \\
  \d{(g,x)}{(h,y)}{(k,z)} & = & \delta_{x,y} \delta_{x,z} \delta_{g,hk}\theta_x(h,k),
  \label{domegag}\\
  \epsilon_{(g,x)} & = &  \delta_{g,e},\nonumber\\
  \s{(h,y)}{(g,x)} 
  & = &  \delta_{gh,e}\delta_{xy,e}\nonumber
  \theta_h(y^{-1},y)^{-1}
  \theta_{y^{-1}}(h,h^{-1})^{-1},
\end{eqnarray}
where
$  \theta_g(x,y) \equiv  \omega(g,x,y)\omega(x,y,g)\omega(x,g,y)^{-1}$.
It is known that $\theta_g(x,y) = \theta_{g^{-1}}(x,y)$ and that
there exists functions $c_g(x)$ on $G$ labeled by $g \in G$ such that
\begin{eqnarray}
  \theta_g(x,y)=c_g(x)c_g(y)c_g(xy)^{-1}. \label{eq:phases}
\end{eqnarray}
It follows that 
\begin{equation}
  \theta_g(x,y)=\theta_g(y,x), \quad 
  \theta_g(x,x^{-1}) \theta_h(x,x^{-1})=\theta_{gh}(x,x^{-1}).
  \label{eq:symmetric}
\end{equation}

Eqs.(\ref{domegag}) imply
\begin{eqnarray}
  C_{(g_1,h_1)(g_2,h_2)\cdots(g_k,h_k)}
  &=& |G|\delta_{g_1,g_2}\delta_{g_1,g_3}\cdots\delta_{g_1,g_k}
  \delta_{h_1\cdots h_k, e} \times \\ \nonumber
  & & \times \prod_{j=1}^{k-1} \theta_{g_1}
  \left(\prod_{\ell=1}^{j}h_\ell,h_{j+1}\right)
  \qquad (\mbox{for\ } k\ge 3),
  \label{Pc123}\\
  \Delta^{(h_1,g_1)(h_2,g_2)\cdots (h_k,g_k)}
  & = & C_{(g_1,h_1)(g_2,h_2)\cdots (g_k,h_k)}. \label{Pd123} 
\end{eqnarray}

We find that $\zcfs_{\D{G}}=|\zdw_{(G;\omega)}|^2$ for
$\omega\equiv 1$ and any commutative finite group $G$.
In fact, $\D{G}$  reduces to the quantum double $D(G)$ and we have 
\begin{eqnarray}
  C_{(g,x)(h,y)(k,z)} & = & \tilde{C}_{ghk} \tilde{\Delta}^{xyz}, 
  \nonumber\\
  \Delta^{(g,x)(h,y)(k,z)} & = & \tilde{\Delta}^{ghk} \tilde{C}_{xyz},
  \label{dgweight}\\
  \s{(h,y)}{(g,x)} 
  & = &  \delta_{gh,e}\delta_{xy,e}.\nonumber
\end{eqnarray}
The quantities with tilde are those for $A=\C[G]$.
Though $D[G]$ is not the tensor product of $\C[G]$'s, the theory
decouples into two sectors and we have
\begin{eqnarray}
  \zcfs_{D(G)}(M) &=& |\zcfs_{\C[G]}(M)|^2.
  \label{twosectors}
\end{eqnarray}
In view of eq.(\ref{dgweight}),
one can imagine these two sectors live 
on the lattice and on the dual lattice, respectively.

For $\omega\not\equiv 1$ theories,  the decoupling
(\ref{dgweight}) does not occur.
However, the partition function 
$\zcfs_{\D{G}}(\Sigma_g \times S^1)$, where $\Sigma_g$ is the closed 
surface with genus $g$,  is the square of that for $\C[G]$:
\begin{equation}
  \zcfs_{D^{\omega}(G)}(\Sigma_g \times S^1) = |G|^{4g}.
  \label{cylinder}
\end{equation}
We can prove this by picking a lattice and  using
(\ref{eq:cocycle}),(\ref{eq:phases}),
(\ref{eq:symmetric}),(\ref{dgweight}), etc.  
Another result we obtain is that each CFS theory has the
factorization property
\begin{equation}
  \zcfs_{\D{G}}(M_1)
  \zcfs_{\D{G}}(M_2) = 
  \zcfs_{\D{G}}(M)
  \zcfs_{\D{G}}(S^3) \mbox{\hspace*{3em}} ( M = M_1 \# M_2).
  \label{factorization}
\end{equation}
We can prove it by calculating 
the weight factor coming from neighborhoods of the
$S^2$ boundaries. 

\begin{figure}[tbhp]
  \begin{center}
    \leavevmode
    \epsfxsize= 200pt 
    \epsfbox{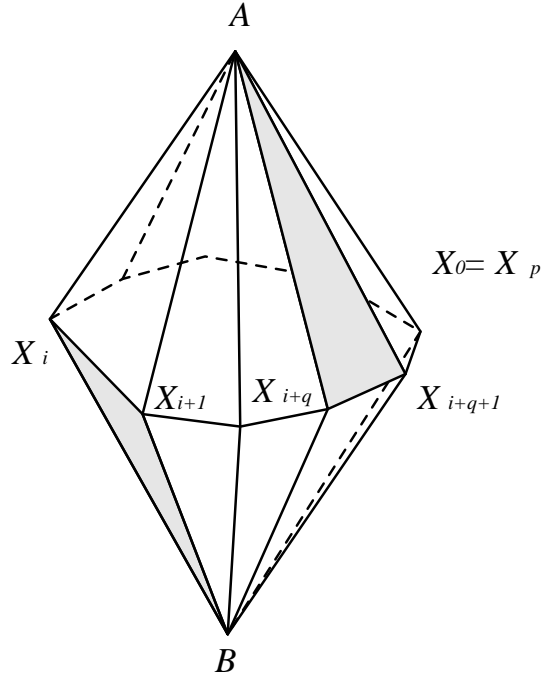}
  \end{center}
  \caption{
    A lattice for $L(p,q)$.  $N_0=2, N_1=p+1, N_2=2p, N_3=p+1$.
    All faces are triangular.
    Edges of two shaded triangles are pair-wisely identified:
    $ BX_iX_{i+1} \sim AX_{i+q}X_{i+q+1}$. As a consequence,
    $A\sim B, X_i \sim X_j$.
    }
  \label{lens}
\end{figure}

Let us explain the case of lens spaces $L(p,q)$ $(p\ge 3)$ in
detail. We take the lattice depicted in fig.~\ref{lens}. 
We obtain the partition function
\begin{eqnarray}
  \zcfs_{\D G}(L(p,q)) 
  & = &  
  |G|^{-2p-2} 
  \prod_{i=0}^{p-1} C_{a_ib_ic_i}C_{x_iy_iz_i}  
  \Delta^{x_0 a_0 x_q a_q x_{2q} a_{2q} x_{3q} a_{3q} 
  \cdots x_{(p-1)q} a_{(p-1)q}}
  \times \nonumber \\
  &   &
  \times \prod_{j=0}^{p-1} \Delta^{b'_j c_{j+1} z_{j+1+q} y'_{j+q}}
  \s{b_j}{b'_j} \s{y_j}{y'_j} \\
  & = &
  |G|^{-2}\sum_{g,h\in G}
  \delta_{h^p,e}\delta_{g^p,e}
  \prod_{i=0}^{p-1}\theta_g(h^i,h)\theta_h(g^i,g).
  \label{dgl}
\end{eqnarray}
Suffices of $x,y,z,a,b,c$ are understood by ${} \bmod p$.
This expression turns out to be true also for $L(0,1)=S^2\times S^1,
L(1,1)=S^3, L(2,1)=\R P^3.$

Let us specialize to the case $G=\Z_N$.
$\Z_N$ has $N$ different 3-cocycles 
$\omega^\ell$ $(\ell=0,1,\ldots,N-1)$ \cite{ms}.  Explicitly,
\begin{eqnarray}
  \omega^\ell_N(x,y,z)
  & = & \exp\left(\frac{2\pi\sqrt{-1} \ell}{N^2}\overline{z}
  (\overline{x}+\overline{y}-\overline{x+ y})\right)
  \label{wln}\\
  &  =  & \left\{ 
  \begin{array}{ll}
    \exp\left(\frac{2\pi\sqrt{-1}\ell}{N}\bar{z}\right) 
                         & \mbox{if $\bar{x}+\bar{y}\geq N$}\\
    1                    
                         & \mbox{otherwise}
  \end{array}
\right. .
\end{eqnarray}
Here, $\bar{x} \in  \{0,1,\ldots,N-1\}$ 
is the representative 
of $x \in \Z/N\Z \simeq \Z_N$:
$\bar{x} \equiv x \bmod{N}$.
In this case, the partition function \reff{dgl} is always a
positive integer independent of $q$:
\begin{eqnarray}
  \frac{\zcfs_{D^{\omega^\ell}(\Z_N)}(L(p,q))}
       {\zcfs_{D^{\omega^\ell}(\Z_N)}(S^3)}
  & = & \sum_{r,s=0}^{m-1}\exp\left(\frac{2p\ell rs}{m^2}\cdot
  2\pi\sqrt{-1}\right)   \label{dznl}\\
  & = & m \times \left| \left\{s \in \{0,\ldots,m-1\} | 2p\ell s \equiv 0
    \bmod m^2   \right\} \right|,
\end{eqnarray}
where $m=(N,p)$ is the greatest common divisor of $N$ and $p$.
We have used 
$ \overline{pz}=0 \Longleftrightarrow \exists r \in
\{0,1,\ldots,m-1\} \mbox{\ s.t.\ } \overline{z}=r N/ m$.

\section{Discussions}
DW theory has been studied elaborately \cite{dw,ac2,fq,br}.
On manifolds $\Sigma_g \times S^1$ and $S^3$,
the DW partition function for a commutative group $G$ 
takes the form 
\begin{equation}
  \zdw_{(G;\omega)}(S^3)=|G|^{-2}, \qquad
  \zdw_{(G;\omega)}(\Sigma_g\times S^1) =  |G|^{2g}.
  \label{eq:dw-3torus}
\end{equation}
For $G=\Z_N$, the partition function on a lens space $L(p,q)$
has an expression
\begin{equation}
  \frac{\zdw_{(\Z_N;\omega^\ell)}(L(p,q))}{\zdw_{(\Z_N;\omega^\ell)}(S^3)}
  =\sum_{r=0}^{m-1}\exp \left(\frac {p\ell nr^2}{m^2}\cdot2\pi\sqrt{-1}\right),
  \label{ca3} 
\end{equation}
where $m = (N,p)$ and $ n \in \{0,\ldots,p-1\}, nq \equiv 1 \bmod p$.

We find that for every $\ell$ and odd $N$,
\begin{eqnarray}
  \zcfs_{D^{\omega^\ell}(\Z_N)}(M) =   |\zdw_{(\Z_N;\omega^\ell)}(M)|^2
  \label{eq:square}
\end{eqnarray}
holds for $M=S^3, \Sigma_g \times S^1$ and $L(p,q)$. 
The relation (\ref{eq:square}) is preserved under the
connected sum provided that the both theories have factorization property.
Therefore the relation (\ref{eq:square}) holds for a wide
variety of manifolds and 
it is suggested that  
$ \zcfs_{D^{\omega^\ell}(\Z_N)}$ and $|\zdw_{(\Z_N;\omega^\ell)}|^2$
for odd $N$ are equivalent as topological invariants.

Eq.(\ref{eq:square}) can be shown as follows.
The proof for $M=S^3, \Sigma_g\times S^1$ is obvious.
For lens spaces, we have
\begin{eqnarray}
|\zdw_{(\Z_N,\omega^\ell)}(L(p,q))|^2
& = & N^{-2} \sum_{r,s=0}^{m-1}
\exp \left( \frac{p\ell n (r^2 -s^2)}{m^2} \cdot 2 \pi \sqrt{-1}\right)\\
& = & N^{-2} \sum_{%
  \begin{scriptsize}
    \begin{array}{l}
      0 \le \alpha \le 2(m-1),\\
      -(m-1) \le \beta \le  (m-1),\\
      \alpha,\beta\in \mbox{\scriptsize\bf Z}, \alpha \equiv \beta \bmod 2
    \end{array}
  \end{scriptsize}}
\exp \left( \frac{p\ell n \alpha \beta}{m^2} \cdot 2 \pi \sqrt{-1}\right)\\
& = & N^{-2} \sum_{\alpha,\beta=0}^{m-1}
\exp \left( \frac{p\ell n \alpha \beta}{m^2} \cdot 2 \pi \sqrt{-1}\right).
\end{eqnarray}
In the second line, we have set $\alpha=r+s, \beta=r-s$. 
In the third line, we have used the fact that the summand 
is invariant under the shift of $\alpha$ or $\beta$ by $m$.
The last expression agrees with (\ref{dznl}) since
$ \{ n \alpha \bmod m | 0 \le \alpha \le m-1\} 
= \{ 2 \alpha \bmod m | 0 \le \alpha \le m-1\} $
due to the equality $(n,m)=(2,m)=1$.

In contrast with the case of odd integer $N$, eq.(\ref{eq:square}) is
not always true for even $N$.
To see this, we consider the theory for  which  $N$ and $\ell$
are both odd.
Then it can be verified that $\zdw_{(\Z_N,\omega^\ell)}(L(2,1))=0$
from eq.(\ref{ca3}).
On the other hand, $\zcfs_{D^{\omega^\ell}(\Z_N)}(L(p,q))$ is always a
positive integer.
Therefore $\zcfs$ and $|\zdw|^2$ cannot be equivalent as
topological invariants. 

We sometimes used the term `topological field theory' for CFS theory above.
It meant, in fact, no more than a set of weights giving rise 
to a topological invariant for closed manifolds.
It is not known whether each CFS theory  has an underlying functor in
Atiyah's axioms \cite{at}.
We expect that CFS theory for a finite-dimensional Hopf algebra $A$ with
the normalization (\ref{normalization})
has the underlying functor.
This is true for $A=\C[G]$ because it is equivalent to DW
theory for $\omega\equiv 1$ and each DW theory has the functor \cite{dw}.
If eq.(\ref{eq:square}) is valid for arbitrary manifolds, the
theory for $A=D^{\omega^\ell}(\Z_{2N+1})$ has an underlying functor,
too. 
It is equivalent to DW theory 
for $G=\Z_N \times \Z_N,$
$\omega((g_1,g_2),(h_1,h_2),(k_1,k_2))=\omega^\ell(g_1,h_1,k_1)
\times \omega^{N-\ell}(g_2,h_2,k_2)$ ,
since 
\begin{equation}
  |\zdw_{(\Z_N;\omega^\ell)}(M)|^2 = 
  \zdw_{(\Z_N;\omega^\ell)}(M) \times
  \zdw_{(\Z_N;\omega^{N-\ell})}(M) =
  \zdw_{(\Z_N\times \Z_N; \omega^\ell \times \omega^{N-\ell})}(M).
\end{equation}

For other theories with $A=\D{G}$,
it is also likely that the functors exist.
Recall that the value of the partition function
$Z(\Sigma_g\times S^1)$ is equal to the
dimension of the Hilbert space ${\cal H}_{\Sigma_g}$ associated
with $\Sigma_g$ in any topological field theories.
In CFS theory,
the partition function $\zcfs_{\D{G}}(\Sigma_g\times S^1)$ is a
positive integer as should be if the functor  exists.
The existence of the functor 
explains the factorization property (\ref{factorization}) 
since $\dim {\cal H}_{S^2} = \zcfs_{\D{G}}(S^2 \times S^1) =1.$

If CFS theory for $A=\D{G}$ has an underlying functor,
it is `irreducible' and cannot be a direct sum 
of other topological field theories  since 
$\dim {\cal H}_{S^2}=1$.
We note that, in the case,  some CFS theories for $\D{\Z_N}$ cannot be 
a tensor product of a number of  DW theories for finite cyclic groups.
Let us assume  that  CFS theory for $A=D^{\omega^1}(\Z_4)$ is a tensor 
product of DW theories.
Due to eqs.(\ref{cylinder}) and (\ref{eq:dw-3torus}) ,
it has to be a product of $\Z_4$ or $\Z_2$ DW theories.
But it can be checked that no combinations of these theories reproduces
$\zcfs_{D^{\omega^1}(\Z_4)}(L(4,1)).$

In DW theory, some pairs of homotopy inequivalent lens spaces
with an identical fundamental group can be distinguished (e.g.
$L(5,1)$ and $L(5,2)$). 
The reversal of the orientation of  a manifold
amounts to complex conjugation of the partition function:
$\zdw(M^*) = \zdw(M)^\ast$ (e.g. $L(3,1)=L(3,2)^\ast$).
Because of the absolute square in (\ref{eq:square}),
$\zcfs_{\D{\Z_{2N+1}}}$ is insensitive of these.
It is desirable to have a Hopf algebra $A$ for which
$\zcfs_A=\zdw_{(\Z_N;\omega)}$. However, we suppose $\zcfs_A$ for
any Hopf algebra $A$ is insensitive of the orientation  because
CFS construction does not refer to the orientation of manifolds.

So far we cannot relate DW theory for non-commutative finite groups 
to CFS theory.
We comment that, for a non-commutative finite group
$G$, $D(G)$ is a Hopf algebra which induces a topological field
theory via CFS construction.  It will be interesting to
investigate such models. 
The decoupling (\ref{dgweight}) does not occur in the models in
general. However, there seems to be no room for twisting since
$\D{G}$ with $\omega\not\equiv 1$ is not always a Hopf algebra.

\subparagraph{Acknowledgments}
The authors thank Prof.~M.~Fukuma for valuable discussions.
They are grateful to Prof.~N.~Sakai for continuous encouragement
and careful reading of a part of the manuscript.
The work by S.H. is supported  by  
Grant-in-Aid for Scientific Research from the Ministry of
Education, Science and Culture.

\end{document}